\documentstyle[aps,preprint]{revtex}\tighten
\def \D {\mbox{D}}
\def \curl {\mbox{curl}\,}
\def \ep {\varepsilon}
\begin{document}

\title{Cosmological dynamics on the brane}

\author{Roy Maartens\thanks{roy.maartens@port.ac.uk}}
\address{~}
\address{Relativity and Cosmology Group, School of
Computer Science and Mathematics, Portsmouth University,
Portsmouth~PO1~2EG, Britain}

\maketitle

\begin{abstract}

In Randall-Sundrum-type brane-world cosmologies, the dynamical
equations on the three-brane differ from the general relativity
equations by terms that carry the effects of imbedding and of the
free gravitational field in the five-dimensional bulk. Instead of
starting from an ansatz for the metric, we derive the covariant
nonlinear dynamical equations for the gravitational and matter
fields on the brane, and then linearize to find the perturbation
equations on the brane. The local energy-momentum corrections are
significant only at very high energies. The imprint on the brane
of the nonlocal gravitational field in the bulk is more subtle,
and we provide a careful decomposition of this effect into
nonlocal energy density, flux and anisotropic stress. The nonlocal
energy density determines the tidal acceleration in the off-brane
direction, and can oppose singularity formation via the
generalized Raychaudhuri equation. Unlike the nonlocal energy
density and flux, the nonlocal anisotropic stress is not
determined by an evolution equation on the brane, reflecting the
fact that brane observers cannot in general make predictions from
initial data. In particular, isotropy of the cosmic microwave
background may no longer guarantee a Friedmann geometry. Adiabatic
density perturbations are coupled to perturbations in the nonlocal
bulk field, and in general the system is not closed on the brane.
But on super-Hubble scales, density perturbations satisfy a
decoupled third-order equation, and can be evaluated by brane
observers. Tensor perturbations on the brane are suppressed by
local bulk effects during inflation, while the nonlocal effects
can serve as a source or a sink. Vorticity on the brane decays as
in general relativity, but nonlocal bulk effects can source the
gravito-magnetic field, so that vector perturbations can be
generated in the absence of vorticity.

\end{abstract}

\pacs{04.50.+h, 98.80.Cq \hfill{hep-th/0004166}}

\section{Introduction}

Einstein's theory of general relativity breaks down at high enough
energies, and is likely to be a limit of a more general theory.
Recent developments in string theory indicate that gravity may be
a truly higher-dimensional theory, becoming effectively
4-dimensional at lower energies. These exciting theoretical
developments need to be accompanied by efforts to test such
higher-dimensional theories against their cosmological and
astrophysical implications. In that spirit, we investigate here a
particular class of models, showing how their dynamical properties
generalize those of Einstein's theory, and discussing the broad
implications of these generalizations for cosmological dynamics.

In many higher-dimensional gravity theories inspired by string
theory, the matter fields are confined to a 3-brane in $1+3+d$
dimensions, while the gravitational field can propagate also in
the $d$ extra dimensions (see, e.g., \cite{gen}). It is not
necessary for the $d$ extra dimensions to be small, or even
finite: recently Randall and Sundrum \cite{rs} have shown that for
$d=1$, gravity can be localized on a single 3-brane even when the
fifth dimension is infinite (see also \cite{old}). An elegant
geometric formulation and generalization of the Randall-Sundrum
scenario has been given by Shiromizu, Maeda and Sasaki \cite{sms}.
The Friedmann equation on the brane in these models is modified by
both high-energy matter terms and a term carrying nonlocal bulk
effects onto the brane. The Friedmann brane models have been
extensively investigated (see, e.g., \cite{bcos,bdel,msm}), and
inflationary scalar perturbations in these models have also been
considered \cite{mwbh}. The models are compatible with
observations subject to reasonable constraints on the parameters.
A broader study of cosmological dynamics, i.e., for induced
metrics more general than the simple Friedmann case, has not been
done. In particular, the analysis of perturbed Friedmann models
also remains to be done. (Considerable work has been done on
perturbations of flat brane metrics; see, e.g.,
\cite{rs,bper,ssm}.)

In this paper, we initiate a study of nonlinear and perturbed
cosmological dynamics in Randall-Sundrum-type brane-world models,
generalizing some important results in general relativity. We find
the bulk corrections to the propagation and constraint equations,
using the covariant Lagrangian approach \cite{ee,m}. This approach
is well-suited to identifying the geometric and physical
quantities that determine inhomogeneity and anisotropy on the
brane, and it is also the basis for a gauge-invariant perturbation
theory \cite{eb}. Our first task is to identify and interpret the
covariant physical content of the bulk effects on the brane. Local
effects lead to quadratic corrections of the density, pressure and
energy flux. The nonlocal effects of the free gravitational field
in the bulk are transmitted by a Weyl projection term, which we
decompose into energy density, energy flux and anisotropic stress
parts. We calculate the gravitational (tidal) and
non-gravitational acceleration of fluid world-lines, finding the
role of the nonlocal energy density in localization of gravity,
and showing how the world-lines have a non-gravitational
acceleration off the brane at high energies. During inflation, the
acceleration is directed towards the brane.

We derive the propagation (`conservation') equations governing the
nonlocal energy density and flux parts; the evolution of the
anisotropic stress part is {\em not} determined on the brane.
These nonlocal terms also enter into crucial dynamical equations,
such as the Raychaudhuri equation and the shear propagation
equation, and can lead to important changes from the general
relativistic case. For example, it is possible via the nonlocal
term to avoid the initial singularity in a nonrotating model
without cosmological constant. Nonlocal effects also mean that
isotropy of the cosmic microwave background (CMB) may no longer
guarantee a Friedmann geometry.

The covariant nonlinear equations lead to a covariant and
gauge-invariant description of perturbations on the brane. We
derive the equations governing adiabatic scalar perturbations,
which are not in general closed on the brane, because of nonlocal
effects. However, on super-Hubble scales, the density
perturbations satisfy a decoupled third-order equation, with an
additional nonlocal degree of freedom, and can therefore be
evaluated by brane observers. Tensor perturbations cannot be
determined by brane observers on any scales. The local bulk
effects tend to enhance tensor perturbations during
non-inflationary expansion and suppress them during inflation.
Nonlocal bulk effects can in principle act either way. Vorticity
on the brane decays as in general relativity, but bulk effects act
as a source for the gravito-magnetic field, and hence vector
perturbations, on the brane.

Our results remain incomplete in one fundamental aspect, i.e., we
do not provide a description of the gravitational field in the
bulk, but confine our investigations to effects that can be
measured by brane-observers. In order to fill this gap, one would
need to study the off-brane derivatives of the curvature, which
are given in general in \cite{sms,ssm}. This is an important topic
for further research.

The 5-dimensional field equations are Einstein's equations, with a
(negative) bulk cosmological constant $\widetilde{\Lambda}$ and
brane energy-momentum as source:
\begin{equation}
\widetilde{G}_{AB} =
\widetilde{\kappa}^2\left[-\widetilde{\Lambda}\widetilde{g}_{AB}
+\delta(\chi)\left\{
-\lambda g_{AB}+T_{AB}\right\}\right]\,. \label{1}
\end{equation}
The tildes denote the bulk (5-dimensional) generalization of
standard general relativity quantities, and
$\widetilde{\kappa}^2= 8\pi/\widetilde{M}_{\rm p}^3$, where
$\widetilde{M}_{\rm p}$ is the fundamental 5-dimensional Planck
mass, which is typically much less than the effective Planck mass
on the brane, $M_{\rm p}=1.2\times 10^{19}$ GeV. The brane is
given by $\chi=0$, so that a natural choice of coordinates is
$x^A=(x^\mu,\chi)$, where $x^\mu=(t,x^i)$ are spacetime
coordinates on the brane. The brane tension is $\lambda$, and
$g_{AB}=\widetilde{g}_{AB}-n_An_B$ is the induced metric on the
brane, with $n_A$ the spacelike unit normal to the brane. Matter
fields confined to the brane make up the brane energy-momentum
tensor $T_{AB}$ (with $T_{AB}n^B=0$).

Although it is usually assumed that the spacetime is exactly
anti-de Sitter in the absence of a brane ($\lambda=0=T_{AB}$),
this is not necessarily the case. The brane-free bulk metric can
be any solution of $\widetilde{G}_{AB} =-
\widetilde{\kappa}^2\widetilde{\Lambda}\widetilde{g}_{AB}$,
including non-conformally flat solutions. When the induced metric
on the brane is Friedmann, then the 5-dimensional
Schwarzschild-anti-de Sitter metric is a solution of
Eq.~(\ref{1})~\cite{msm}. However, more general bulk metrics are
in principle possible.

The field equations induced on the brane are derived via an
elegant geometric approach by Shiromizu et al. \cite{sms}, using
the Gauss-Codazzi equations, matching conditions and $Z_2$
symmetry. The result is a modification of the standard Einstein
equations, with the new terms carrying bulk effects onto the
brane:\footnote{ For clarity and consistency, we have changed the
notation of \cite{sms}.}
\begin{equation}
G_{\mu\nu}=-\Lambda g_{\mu\nu}+\kappa^2
T_{\mu\nu}+\widetilde{\kappa}^4S_{\mu\nu} - {\cal E}_{\mu\nu}\,,
\label{2}
\end{equation}
where $\kappa^2=8\pi/M_{\rm p}^2$. The energy scales are related
to each other via
\begin{equation}
\lambda=6{\kappa^2\over\widetilde\kappa^4} \,, ~~ \Lambda =
{4\pi\over \widetilde{M}_{\rm p}^3}\left[\widetilde{\Lambda}+
\left({4\pi\over 3\widetilde{M}_{\rm
p}^{\,3}}\right)\lambda^2\right]\,. \label{3}
\end{equation}

The bulk corrections to the Einstein equations on the brane are of
two forms: firstly, the matter fields contribute local quadratic
energy-momentum corrections via the tensor $S_{\mu\nu}$, and
secondly, there are nonlocal effects from the free gravitational
field in the bulk, transmitted via the projection ${\cal
E}_{\mu\nu}$ of the bulk Weyl tensor. The matter corrections are
given by
\begin{equation}
S_{\mu\nu}={\textstyle{1\over12}}T_\alpha{}^\alpha T_{\mu\nu}
-{\textstyle{1\over4}}T_{\mu\alpha}T^\alpha{}_\nu+
{\textstyle{1\over24}}g_{\mu\nu} \left[3 T_{\alpha\beta}
T^{\alpha\beta}-\left(T_\alpha{}^\alpha\right)^2 \right]\,.
\label{3'}
\end{equation}
The projection of the bulk Weyl tensor is\footnote{ This
projection is called the `electric' part of the Weyl tensor in
\cite{sms}, but the term is potentially misleading, since the
electric part is associated with projection on a timelike
vector~\cite{ee,s}, and $n^A$ is spacelike. ${\cal E}_{\mu\nu}$
should not be confused with the electric part of the {\em brane}
Weyl tensor, $E_{\mu\nu}$, defined below.}
\begin{equation}
{\cal E}_{AB}=\widetilde{C}_{ACBD}n^C n^D\,,
\label{4}
\end{equation}
which is symmetric\footnote{ (Square) round brackets enclosing
indices denote (anti-)symmetrization.} and traceless (${\cal
E}_{[AB]}=0={\cal E}_A{}^A$) and without components orthogonal to
the brane, so that ${\cal E}_{AB}n^B=0$ and ${\cal E}_{AB}\to
{\cal E}_{\mu\nu}\delta_A{}^\mu\delta_B{}^\nu$ as $\chi\to 0$. The
Weyl tensor $\widetilde{C}_{ABCD}$ represents the free, nonlocal
gravitational field in the bulk. The local part of the bulk
gravitational field is the Einstein tensor $\widetilde{G}_{AB}$,
which is determined via the bulk field equations (\ref{1}). Thus
${\cal E}_{\mu\nu}$ {\em transmits nonlocal gravitational degrees
of freedom from the bulk to the brane, including tidal (or
Coulomb), gravito-magnetic and transverse traceless (gravitational
wave) effects.}

\section{Covariant decomposition of bulk effects}

We now provide a covariant decomposition of the bulk correction
tensors given by Shiromizu et al. \cite{sms}.

\subsection{Local bulk effects}

For {\em any} matter fields (scalar fields, perfect fluids,
kinetic gases, dissipative fluids, etc.), including a combination
of different fields, the general form of the brane energy-momentum
tensor can be covariantly given as
\begin{equation}
T_{\mu\nu}=\rho u_\mu
u_\nu+ph_{\mu\nu}+\pi_{\mu\nu}+2q_{(\mu}u_{\nu)}\,.
 \label{3''}
\end{equation}
The decomposition is irreducible for any chosen 4-velocity
$u^\mu$. Here $\rho$ and $p$ are the energy density and isotropic
pressure, and $h_{\mu\nu}=g_{\mu\nu}+u_\mu u_\nu$ projects
orthogonal to $u^\mu$. The energy flux obeys
$q_{\mu}=q_{\langle\mu\rangle}$, and the anisotropic stress obeys
$\pi_{\mu\nu}=\pi_{\langle\mu\nu\rangle}$, where angle brackets
denote the projected, symmetric and tracefree part:
\[
V_{\langle\mu\rangle}=h_\mu{}^\nu V_\nu\,,~~
W_{\langle\mu\nu\rangle}=\left[h_{(\mu}{}^\alpha h_{\nu)}{}^\beta-
{\textstyle{1\over3}}h^{\alpha\beta}h_{\mu\nu}\right]W_{\alpha\beta}\,.
\]

Equations (\ref{3'}) and (\ref{3''}) imply the irreducible
decomposition
\begin{eqnarray}
S_{\mu\nu}&=&{\textstyle{1\over24}}\left[2\rho^2-3\pi_{\alpha\beta}
\pi^{\alpha\beta}\right]u_\mu u_\nu
+{\textstyle{1\over24}}\left[2\rho^2+4\rho p+\pi_{\alpha\beta}
\pi^{\alpha\beta}-4q_\alpha q^\alpha\right]h_{\mu\nu} \nonumber\\
&&{}-
{\textstyle{1\over12}}(\rho+2p)\pi_{\mu\nu}+\pi_{\alpha\langle\mu}
\pi_{\nu\rangle}{}^\alpha+q_{\langle\mu}q_{\nu\rangle}+
{\textstyle{1\over3}}\rho q_{(\mu}u_{\nu)}- {\textstyle{1\over12}}
q^\alpha \pi_{\alpha(\mu}u_{\nu)} \,. \label{3'''}
\end{eqnarray}
For a perfect fluid or minimally-coupled scalar field
\[
S_{\mu\nu}={\textstyle{1\over12}}\rho^2
u_\mu u_\nu
+{\textstyle{1\over12}}\rho\left(\rho+2 p\right)h_{\mu\nu}\,,
\]
in agreement with \cite{sms}. The quadratic energy-momentum
corrections to standard general relativity will be significant for
$\widetilde{\kappa}^4\rho^2 \gtrsim \kappa^2\rho$, i.e., in the
high-energy regime
\[
\rho\gtrsim \lambda \sim
\left({\widetilde{M}_{\rm p}\over M_{\rm p}}\right)^2
\widetilde{M}_{\rm p}^{\,4}\,.
\]

\subsection{Nonlocal bulk effects}

The symmetry properties of ${\cal E}_{\mu\nu}$ imply that in
general we can decompose it irreducibly with respect to a chosen
4-velocity field $u^\mu$ as
\begin{equation}
{\cal
E}_{\mu\nu}=-\left({\widetilde{\kappa}\over\kappa}\right)^4\left[{\cal
U}\left(u_\mu u_\nu+{\textstyle {1\over3}} h_{\mu\nu}\right)+{\cal
P}_{\mu\nu}+2{\cal Q}_{(\mu}u_{\nu)}\right]\,. \label{6}
\end{equation}
The factor $(\widetilde\kappa/\kappa)^4$ is introduced for
dimensional reasons. Here
\[
{\cal U}=-\left({{\kappa}\over\widetilde\kappa}\right)^4
{\cal E}_{\mu\nu}u^\mu u^\nu
\]
is an effective nonlocal energy density on the brane, arising from
the free gravitational field in the bulk. This nonlocal energy
density need not be positive (see below). There is an effective
nonlocal anisotropic stress
\[
{\cal P}_{\mu\nu}=-\left({{\kappa}\over\widetilde\kappa}\right)^4
{\cal E}_{\langle\mu\nu\rangle}
\]
on the brane, arising from the free gravitational field in the
bulk, and
\[
{\cal Q}_\mu =\left({{\kappa}\over\widetilde\kappa}\right)^4
{\cal E}_{\langle\mu\rangle\nu}u^\nu
\]
is an effective nonlocal energy flux on the brane, arising from
the free gravitational field in the bulk.

If the induced metric on the brane is flat, and the bulk is
anti-de Sitter, as in the original Randall-Sundrum scenario
\cite{rs}, then ${\cal E}_{\mu\nu}=0$. Treating this as a
background, it follows that tensor (transverse traceless)
perturbations on the brane arising from nonlocal bulk degrees of
freedom are given by
\begin{equation}\label{6a}
{\cal U}=0={\cal Q}_{\mu}\,,~\D^\nu{\cal P}_{\mu\nu}=0\,,
\end{equation}
where $\D_\mu$ is the totally projected part of the brane
covariant derivative:
\[
\D_\mu F^{\alpha\cdots}{}{}_{\cdots\beta}=h_\mu{}^\nu
h^\alpha{}_\gamma \cdots h_\beta{}^\delta \nabla_\nu
F^{\gamma\cdots}{}{}_{\cdots\delta}\,.
\]
Equation (\ref{6a}) provides a covariant characterization of brane
tensor perturbations on an anti-de Sitter background.

In cosmology, the background induced metric is not flat, but a
spatially homogeneous and isotropic Friedmann model, for which
\begin{equation}\label{6c}
\D_\mu{\cal U}={\cal Q}_{\mu}={\cal P}_{\mu\nu}=0\,.
\end{equation}
Thus for a perturbed Friedmann model, the nonlocal bulk effects
are covariantly and gauge-invariantly described by the first-order
quantities $\D_\mu{\cal U}$, ${\cal Q}_{\mu}$, ${\cal
P}_{\mu\nu}$. Since ${\cal U}\neq0$ in general in the Friedmann
background \cite{bcos,bdel}, it follows that for tensor
perturbations on the brane
\begin{equation}\label{6a'}
\D_\mu{\cal U}=0={\cal Q}_{\mu}\,,~\D^\nu{\cal P}_{\mu\nu}=0\,.
\end{equation}
Scalar perturbations (Coulomb-like bulk effects) will be
characterized by
\begin{equation}\label{6b'}
{\cal Q}_{\mu}=\D_\mu{\cal Q}\,,~{\cal P}_{\mu\nu}=\D_{\langle\mu}
\D_{\nu\rangle}{\cal P} \,,
\end{equation}
for some scalars ${\cal Q}$ and ${\cal P}$, while for vector
perturbations (gravito-magnetic-like bulk effects)
\begin{equation}\label{6c'}
\D^\mu{\cal Q}_{\mu}=0\,,~
{\cal P}_{\mu\nu}=\D_{\langle\mu}{\cal P}_{\nu\rangle} \,,~
\D^\mu{\cal P}_{\mu}=0\,.
\end{equation}

\subsection{Gravitational and non-gravitational acceleration}

In order to find the role of bulk effects in tidal acceleration on
the brane, we start from the relation
\[
{\cal E}_{\mu\nu}{u}^\mu u^\nu=\lim_{\chi\to 0}
\widetilde{\cal C}_{ABCD}\widetilde{u}^A n^B
\widetilde{u}^C n^D\,,
\]
where $\widetilde{u}^A$ is an extension off the brane of the
4-velocity (with $\widetilde{u}^An_A=0$). The tidal acceleration
in the $n^A$ direction measured by comoving observers is
$-n_A\widetilde{R}^{\,A}{}_{BCD}\widetilde{u}^Bn^C
\widetilde{u}^D$. Now
\[
\widetilde{R}_{ABCD}= \widetilde{C}_{ABCD}+{\textstyle{2\over3}}
\left\{\widetilde{g}_{A[C}\widetilde{R}_{D]B}+
\widetilde{g}_{B[D}\widetilde{R}_{C]A}\right\}-{\textstyle{1\over6}}
\widetilde{R}\widetilde{g}_{A[C}\widetilde{g}_{D]B}\,,
\]
so that by the field equation (\ref{1}) (and recalling
that $T_{AB}n^B=0$),
\[
-\widetilde{R}_{ABCD}n^A \widetilde{u}^Bn^C
\widetilde{u}^D=-{\cal E}_{AB}\widetilde{u}^A\widetilde{u}^B+
{\textstyle{1\over6}}\widetilde{\kappa}^2\widetilde{\Lambda}\,.
\]
Thus {\em the comoving tidal acceleration on the brane, in the
off-brane direction, is}
\begin{equation}\label{6b}
\left({\widetilde\kappa\over\kappa}\right)^4{\cal U}
+{\textstyle{1\over6}}\widetilde{\kappa}^2\widetilde{\Lambda}\,.
\end{equation}
Since $\widetilde{\Lambda}$ is negative, it contributes to
acceleration towards the brane. This reflects the confining role
of the negative bulk cosmological constant on the gravitational
field in the warped metric models of Randall-Sundrum type.
Equation~(\ref{6b}) shows that {\em localization of the
gravitational field near the brane is enhanced by negative ${\cal
U}$, while positive ${\cal U}$ acts against confinement,} by
contributing to tidal acceleration away from the brane. This
picture is consistent with a Newtonian interpretation, where the
gravitational field carries negative energy density.

The non-gravitational acceleration of fluid world-lines on the
brane is
\[
\widetilde{A}^A=\widetilde{u}^B\widetilde{\nabla}_B
\widetilde{u}^A~\mbox{ at }~ \chi=0\,.
\]
Locally, near the brane, the metric may be written as \cite{sms}
\[
d\tilde{s}^2=d\chi^2+g_{\mu\nu}(x^\alpha,\chi)dx^\mu dx^\nu\,,
\]
so that
\[
\widetilde{\Gamma}^A{}_{\mu\nu}(x,0)=\Gamma^\alpha{}_{\mu\nu}(x)
\delta_\alpha{}^A-{\textstyle{1\over2}}g_{\mu\nu,\chi}(x,0)
\delta_\chi{}^A\,.
\]
This allows us to express the 5-dimensional acceleration
$\widetilde{A}^A$ in terms of the 4-dimensional acceleration
$A^\mu=u^\nu\nabla_\nu u^\mu$ on the brane. First, we use the
covariant form $g_{\mu\nu,\chi}={\cal L}_{\bf n}g_{\mu\nu}$, where
${\cal L}_{\bf n}$ is the Lie derivative along $n^A$. Then we use
the expression for the extrinsic curvature of the brane \cite{sms}
\[
K^+_{\mu\nu}={\textstyle{1\over2}}\lim_{\chi\to 0^+}
{\cal L}_{\bf n}g_{\mu\nu}\,,
\]
which leads to
\begin{equation}\label{acc}
\widetilde{A}^A(x,0^+)=
A^\mu(x)\delta_\mu{}^A-K^+_{\mu\nu}(x)u^\mu(x) u^\nu(x) n^A(x,0^+)\,,
\end{equation}
on the brane. The extrinsic curvature is given in terms of the
brane tension and energy-momentum by \cite{sms}
\[
K^+_{\mu\nu}=-{\textstyle{1\over6}}\widetilde{\kappa}^2\left[
\lambda g_{\mu\nu}+3T_{\mu\nu}+(\rho-3p)g_{\mu\nu}\right]\,.
\]
Substituting in Eq. (\ref{acc}), we find
\begin{equation}\label{acc'}
\widetilde{A}^A(x,0^+)=
A^\mu(x)\delta_\mu{}^A+{\textstyle{1\over6}}\widetilde{\kappa}^2\left[
2\rho(x)+3p(x)-\lambda\right]n^A(x,0^+)\,.
\end{equation}

It follows that $n_A\widetilde{A}^A$ is nonzero on the brane,
i.e., there is {\em a non-gravitational acceleration of fluid
world-lines orthogonal to the brane}. The direction depends on the
sign of $2\rho+3p-\lambda$ : {\em for $2\rho+3p-\lambda>0$, the
transverse acceleration is away from the brane, while for
$2\rho+3p-\lambda<0$, it is towards the brane.}  If the pressure
is positive, then at high energies $2\rho+3p-\lambda>0$, so that
either the brane must accelerate, or there must be a
non-gravitational mechanism for keeping matter on the brane.

Inflationary expansion, when pressure is negative, can change this
situation. Inflation on the brane is characterized by~\cite{mwbh}
\begin{equation}\label{inf}
p<-\left({\lambda+2\rho\over\lambda+\rho}\right){\rho\over3}\,,
\end{equation}
and this condition implies that $2\rho+3p-\lambda<0$. Thus during
inflation on the brane, the transverse acceleration is towards the
brane; {\em inflation acts as a non-gravitational mechanism
keeping matter on the brane.}

\subsection{Effective total energy-momentum tensor}

All the bulk corrections may be consolidated into effective total
energy density, pressure, anisotropic stress and energy flux, as
follows. The modified Einstein equations take the standard
Einstein form with a redefined energy-momentum tensor:
\begin{equation}
G_{\mu\nu}=-\Lambda g_{\mu\nu}+\kappa^2 T^{\rm tot}_{\mu\nu}\,,
\label{6'}
\end{equation}
where
\begin{equation}
T^{\rm tot}_{\mu\nu}= T_{\mu\nu}+{\widetilde{\kappa}^{4}\over
\kappa^2}S_{\mu\nu}- {1\over\kappa^2}{\cal E}_{\mu\nu}\,.
\label{6''}
\end{equation}
Then
\begin{eqnarray}
\rho^{\rm tot} &=& \rho+{\widetilde{\kappa}^{4}\over
\kappa^6}\left[{\kappa^4\over24}\left(2\rho^2 -3
\pi_{\mu\nu}\pi^{\mu\nu}\right) +{\cal U}\right] \label{a}\\
p^{\rm tot} &=& p+ {\widetilde{\kappa}^{4}\over
\kappa^6}\left[{\kappa^4\over24}\left(2\rho^2+4\rho p+
\pi_{\mu\nu}\pi^{\mu\nu}-4q_\mu q^\mu\right) +{\textstyle{1\over
3}}{\cal U}\right] \label{b}\\ \pi^{\rm tot}_{\mu\nu} &=&
\pi_{\mu\nu}+ {\widetilde{\kappa}^{4}\over
\kappa^6}\left[{\kappa^4\over12}\left\{-(\rho+3p)\pi_{\mu\nu}+
\pi_{\alpha\langle\mu}\pi_{\nu\rangle}{}^\alpha+q_{\langle\mu}q_
{\nu\rangle}\right\} +{\cal P}_{\mu\nu}\right]\label{c}\\ q^{\rm
tot}_\mu &=&q_\mu+ {\widetilde{\kappa}^{4}\over
\kappa^6}\left[{\kappa^4\over24}\left(4\rho
q_\mu-\pi_{\mu\nu}q^\nu\right)+ {\cal Q}_\mu\right] \,.\label{d}
\end{eqnarray}
(Note that $\widetilde{\kappa}^{4}/\kappa^6$ is dimensionless.)

These general expressions simplify in the case of a perfect fluid
(or minimally coupled scalar field, or isotropic one-particle
distribution function), i.e., for $q_\mu=0=\pi_{\mu\nu}$. However,
the total energy flux and anisotropic stress do not vanish in this
case in general:
\[
q^{\rm tot}_\mu ={\widetilde{\kappa}^{4}\over \kappa^6}{\cal
Q}_\mu\,,~~ \pi^{\rm tot}_{\mu\nu} ={\widetilde{\kappa}^{4}\over
\kappa^6}{\cal P}_{\mu\nu}\,.
\]
Thus {\em nonlocal bulk effects can contribute to effective
imperfect fluid terms even when the matter on the brane has
perfect fluid form.}

\section{Local and nonlocal conservation equations}

As a consequence of the form of the bulk energy-momentum tensor in
Eq. (\ref{1}) and of $Z_2$ symmetry, it follows \cite{sms} that
the brane energy-momentum tensor separately satisfies the
conservation equations, i.e.,
\begin{equation}\label{5'}
\nabla^\nu T_{\mu\nu}=0 \,.
\end{equation}
Then the Bianchi identities on the brane imply that the projected
Weyl tensor obeys the constraint
\begin{equation}
\nabla^\mu{\cal E}_{\mu\nu}=\widetilde{\kappa}^4\nabla^\mu
S_{\mu\nu}\,, \label{5}
\end{equation}
which shows that its longitudinal part is sourced by quadratic
energy-momentum terms, including spatial gradients and time
derivatives. Thus evolution and inhomogeneity in the matter fields
generates nonlocal Coulomb-like gravitational effects in the bulk,
which `backreact' on the brane.

The brane energy-momentum tensor {\em and} the consolidated
effective energy-momentum tensor are {\em both} conserved
separately. Conservation of $T_{\mu\nu}$ gives the standard
general relativity energy and momentum conservation
equations~\cite{cov}
\begin{eqnarray}
&&\dot{\rho}+\Theta(\rho+p)+\D^\mu q_\mu+2A^\mu
q_\mu+\sigma^{\mu\nu}\pi_{\mu \nu}=0\,,\label{c1}\\ &&
\dot{q}_{\langle\mu\rangle}+{\textstyle{4\over3}}\Theta
q_\mu+\D_\mu p+(\rho+p)A_\mu + \D^\nu
\pi_{\mu\nu}+A^\nu\pi_{\mu\nu}+\sigma_{\mu\nu}q^\nu- [\omega,
q]_\mu=0\,.\label{c2}
\end{eqnarray}
A dot denotes $u^\nu\nabla_\nu$, $\Theta=\D^\mu u_\mu$ is the
volume expansion rate of the $u^\mu$ congruence,
$A_\mu=\dot{u}_\mu =A_{\langle\mu\rangle}$ is its 4-acceleration,
$\sigma_{\mu\nu}=\D_{\langle\mu}u_{\nu\rangle}$ is its shear rate,
and $\omega_\mu=-{1\over2}\curl u_\mu=\omega_{\langle\mu\rangle}$
is its vorticity rate. The covariant spatial curl is given
by~\cite{m}
\[
\curl V_\mu=\ep_{\mu\alpha\beta}\D^\alpha V^\beta\,,~~ \curl W
_{\mu\nu}=\ep_{\alpha\beta(\mu}\D^\alpha W^\beta{}_{\nu)}\,,
\]
where $\ep_{\mu\nu\sigma}$ is the projected alternating tensor.
The covariant cross-product is
\[
[V,Y]_\mu=\ep_{\mu\alpha\beta}V^\alpha Y^\beta\,.
\]

The conservation of $T^{\rm tot}_{\mu\nu}$ gives, upon using Eqs.
(\ref{a})--(\ref{c2}), a propagation equation for the nonlocal
energy density ${\cal U}$:
\begin{eqnarray}
&& \dot{\cal U}+{\textstyle{4\over3}}\Theta{\cal U}+\D^\mu{\cal
Q}_\mu+2A^\mu{\cal Q}_\mu+\sigma^{\mu\nu}{\cal
P}_{\mu\nu}\nonumber\\ &&~~{}={\textstyle{1\over24}}\kappa^4\left[
6\pi^{\mu\nu}\dot{\pi}_{\mu\nu}+6(\rho+p)\sigma^{\mu\nu}
\pi_{\mu\nu}+2\Theta \left(2q^\mu
q_\mu-\pi^{\mu\nu}\pi_{\mu\nu}\right)\right.\nonumber\\
&&~~~\left.{} +2A^\mu
q^\nu\pi_{\mu\nu}+4q^\mu\D_\mu\rho+q^\mu\D^\nu\pi_{\mu\nu}
+\pi^{\mu\nu}\D_\mu q_\nu -
2\sigma^{\mu\nu}\pi_{\alpha\mu}\pi_\nu{}^\alpha-
2\sigma^{\mu\nu}q_\mu q_\nu \right]\,, \label{c1'}
\end{eqnarray}
and a propagation equation for the nonlocal energy flux ${\cal
Q}_\mu$:
\begin{eqnarray}
&& \dot{\cal
Q}_{\langle\mu\rangle}+{\textstyle{4\over3}}\Theta{\cal Q}_\mu
+{\textstyle{1\over3}}\D_\mu{\cal U}+{\textstyle{4\over3}}{\cal
U}A_\mu +\D^\nu{\cal P}_{\mu\nu}+A^\nu{\cal
P}_{\mu\nu}+\sigma_{\mu\nu}{\cal Q}^\nu-[\omega,{\cal Q}]_\mu
\nonumber\\ &&~~{}={\textstyle{1\over24}} \kappa^4 \left[
-4(\rho+p)\D_\mu
\rho+q^\nu\dot{\pi}_{\langle\mu\nu\rangle}+\pi_\mu{}^\nu
\D_\nu(2\rho+5p)+6(\rho+p)\D^\nu\pi_{\mu\nu}\right.\nonumber\\
&&~~~\left.{}-{\textstyle{2\over3}}\pi^{\alpha\beta}
\left(\D_\mu\pi_{\alpha
\beta}+3\D_\alpha\pi_{\beta\mu}\right)-3\pi_{\mu\alpha}\D_\beta
\pi^{\alpha\beta}+{\textstyle{28\over3}}q^\nu\D_\mu q_\nu
\right.\nonumber\\ &&~~~\left.{}+4\rho
A^\nu\pi_{\mu\nu}-3\pi_{\mu\alpha}A_\beta\pi^{\alpha\beta}
+{\textstyle{8\over3}}A_\mu\pi^{\alpha\beta}\pi_{\alpha\beta}
-\pi_{\mu\alpha}\sigma^{\alpha\beta}q_\beta+\sigma_{\mu\alpha}
\pi^{\alpha\beta}q_\beta+ \pi_{\mu\nu}[\omega, q]^\nu
\right.\nonumber\\&&~~~\left.{}-\ep_{\mu\alpha\beta}\omega^\alpha
\pi^{\beta\nu}q_\nu+4(\rho+p)\Theta q_\mu+ 6q_\mu A^\nu q_\nu
+{\textstyle{14\over3}}A_\mu q^\nu q_\nu+4q_\nu
\sigma^{\alpha\beta} \pi_{\alpha\beta}\right]\,.\label{c2'}
\end{eqnarray}
These equations may be thought of as conservation equations on the
brane for the nonlocal energy density and energy flux due to the
free gravitational field in the bulk. In general, the 4
independent equations determine 4 of the 9 independent components
of ${\cal E}_{\mu\nu}$ on the brane. What is missing, is an
evolution equation for ${\cal P}_{\mu\nu}$ (which has up to 5
independent components). Thus in general, the projection of the
5-dimensional field equations onto the brane does not lead to a
closed system. Nor could we expect this to be the case, since
there are bulk degrees of freedom whose impact on the brane cannot
be predicted by brane observers. These degrees of freedom could
arise from propagating gravity waves in the bulk, possibly
generated by inhomogeneity on the brane itself. The point is that
waves which penetrate the 5th dimension are governed by off-brane
bulk dynamical equations. Our decomposition of ${\cal E}_{\mu\nu}$
has shown that {\em the evolution of the nonlocal energy density
and flux (associated with the scalar and vector parts of ${\cal
E}_{\mu\nu}$) is determined on the brane, while the evolution of
the nonlocal anisotropic stress (associated with the tensor part
of ${\cal E}_{\mu\nu}$) is not.}

If the nonlocal anisotropic stress contribution from the bulk
field vanishes, i.e., if
\[
{\cal P}_{\mu\nu}=0\,,
\]
then the evolution of ${\cal E}_{\mu\nu}$ is determined by Eqs.
(\ref{c1'}) and (\ref{c2'}). A special case of this arises when
the induced metric on the brane is Friedmann, i.e., when Eq.
(\ref{6c}) holds. Then ${\cal E}_{\mu\nu}$ has only 1 independent
component, ${\cal U}$, and it is determined by Eq. (\ref{c1'})
(see below), with Eq. (\ref{c2'}) reducing to $0=0$. Another case
when the equations close on the brane is when the brane is static
(see \cite{dmpr}).

In the perfect fluid case, the conservation equations
(\ref{c1})--(\ref{c2'}) reduce to:
\begin{eqnarray}
&&\dot{\rho}+\Theta(\rho+p)=0\,,\label{pc1}\\ && \D_\mu
p+(\rho+p)A_\mu =0\,.\label{pc2}
\end{eqnarray}
For a minimally-coupled scalar field,
$\rho={1\over2}\nabla_\mu\varphi \nabla^\mu\varphi+V(\varphi)$ and
$p={1\over2}\nabla_\mu\varphi \nabla^\mu\varphi-V(\varphi)$. In
the adiabatic case, Eq. (\ref{pc2}) gives
\begin{equation}\label{ad}
A_\mu=-{c_{\rm s}^2\over\rho+p}\,\D_\mu\rho\,,~~
c_{\rm s}^2={\dot{p}\over\dot{\rho}}\,.
\end{equation}

The nonlocal conservation equations (\ref{c1'}) and (\ref{c2'})
reduce to
\begin{eqnarray}
&& \dot{\cal
U}+{\textstyle{4\over3}}\Theta{\cal U}+\D^\mu{\cal
Q}_\mu+2A^\mu{\cal Q}_\mu+\sigma^{\mu\nu}{\cal P}_{\mu\nu}=0\,,
\label{pc1'}\\&& \dot{\cal
Q}_{\langle\mu\rangle}+{\textstyle{4\over3}}\Theta{\cal Q}_\mu
+{\textstyle{1\over3}}\D_\mu{\cal U}+{\textstyle{4\over3}}{\cal
U}A_\mu +\D^\nu{\cal P}_{\mu\nu}+A^\nu{\cal
P}_{\mu\nu}+\sigma_{\mu\nu}{\cal Q}^\nu-[\omega,{\cal Q}]_\mu
\nonumber\\&&~~{}
=-{\textstyle{1\over6}} \kappa^4 (\rho+p)\D_\mu
\rho\,.\label{pc2'}
\end{eqnarray}

Equation (\ref{pc2'}) shows that {\em if ${\cal E}_{\mu\nu}=0$ and
the brane energy-momentum tensor has perfect fluid form, then the
density $\rho$ must be homogeneous}~\cite{sms}. The converse does
not hold, i.e., homogeneous density does {\em not} in general
imply vanishing ${\cal E}_{\mu\nu}$. This is readily apparent from
Eq.~(\ref{pc2'}). A simple example is provided by the Friedmann
case: Eq. (\ref{pc2'}) is trivially satisfied, while Eq.
(\ref{pc1'}) gives the `dark radiation' solution
\begin{equation}\label{dr}
{\cal U}={\cal U}_o\left({a_o\over a}\right)^4\,.
\end{equation}
A simple generalization of the Friedmann case is the purely
Coulomb-like case, ${\cal Q}_\mu=0={\cal P}_{\mu\nu}$. Equation
(\ref{dr}) still holds, but with $a$ an average scale factor,
which is in general inhomogeneous. Equation (\ref{pc2'}) reduces
to a constraint on the acceleration. Local momentum conservation
already provides the constraint in Eq. (\ref{ad}). It follows that
in the purely Coulomb-like case, the spatial gradient of the
nonlocal energy density must be proportional to that of the local
energy density:
\[
\D_\mu{\cal U}=\left[{8c_{\rm s}^2{\cal U}-\kappa^4 (\rho+p)^2
\over 2(\rho+p)}\right]\D_\mu\rho\,.
\]

Linearization about a Friedmann background does not change Eqs.
(\ref{pc1}) and (\ref{pc2}), but Eqs. (\ref{pc1'}) and
(\ref{pc2'}) lead to
\begin{eqnarray}
&& \dot{\cal U}+{\textstyle{4\over3}}\Theta{\cal U}+\D^\mu{\cal
Q}_\mu  =0
\,, \label{lc1'}\\&& \dot{\cal
Q}_{\mu}+4H{\cal Q}_\mu
+{\textstyle{1\over3}}\D_\mu{\cal U}+{\textstyle{4\over3}}{\cal
U}A_\mu +\D^\nu{\cal P}_{\mu\nu}
=-{\textstyle{1\over6}} \kappa^4(\rho+p)
\D_\mu \rho
\,,\label{lc2'}
\end{eqnarray}
where $H$ is the Hubble rate in the background. The nonlocal
tensor mode, which satisfies $\D^\nu{\cal P}_{\mu\nu}=0 \neq {\cal
P}_{\mu\nu}$, does not enter the nonlocal conservation equations.

\section{Propagation and constraint equations}

Equations (\ref{c1})--(\ref{c2'}) are propagation equations for
the local and nonlocal energy density and flux. The remaining
covariant equations on the brane are the propagation and
constraint equations for the kinematic quantities and the free
gravitational field on the brane. The kinematic quantities govern
the relative motion of neighboring fundamental world-lines, and
describe the universal expansion and its local anisotropies. The
locally free gravitational field {\em on the brane} is given by
the brane Weyl tensor $C_{\mu\nu\alpha\beta}$. This splits
irreducibly for a given $u^\mu$ into the gravito-electric and
gravito-magnetic fields on the brane:
\[
E_{\mu\nu}=C_{\mu\alpha\nu\beta}u^\alpha u^\beta
=E_{\langle\mu\nu\rangle }\,,~~
H_{\mu\nu}={\textstyle{1\over2}}\ep_{\mu\alpha \beta}
C^{\alpha\beta}{}{}_{\nu\gamma}u^\gamma=H_{\langle\mu\nu\rangle}
\,,
\]
where $E_{\mu\nu}$ must not be confused with ${\cal E}_{\mu\nu}$.
The Ricci identity for $u^\mu$ and the Bianchi identities
$\nabla^\beta C_{\mu\nu\alpha\beta} =
\nabla_{[\mu}(-R_{\nu]\alpha} + {1\over6}Rg_{\nu]\alpha})$ produce
the fundamental evolution and constraint equations governing the
above covariant quantities \cite{ee}. Einstein's equations are
incorporated via the algebraic replacement of the Ricci tensor
$R_{\mu\nu}$ by the effective total energy-momentum tensor,
according to Eq. (\ref{6'}). These are derived directly from the
standard general relativity versions (see, e.g., \cite{cov}) by
simply replacing the energy-momentum tensor terms $\rho,\dots$ by
$\rho^{\rm tot},\dots$. The result for a general imperfect fluid
is given in Appendix A.

\subsection{Nonlinear equations}

For a perfect fluid or minimally-coupled scalar field, the
equations in Appendix A reduce to the following.\\
\newpage
\noindent Expansion propagation (generalized Raychaudhuri equation):
\begin{eqnarray}
&&\dot{\Theta}+{\textstyle{1\over3}}\Theta^2+\sigma_{\mu\nu}
\sigma^{\mu\nu}
-2\omega_\mu\omega^\mu -{\rm D}^\mu A_\mu+A_\mu
A^\mu+{\textstyle{1\over2}}\kappa^2(\rho + 3p) -\Lambda
\nonumber\\&&~~{}=
-{1\over12}\left({\widetilde{\kappa}\over\kappa}\right)^4 \left[
{\kappa}^4 \rho(2\rho+3p)
 +12{\cal U}\right]\,. \label{pr}
\end{eqnarray}
Vorticity propagation:
\begin{equation}
\dot{\omega}_{\langle \mu\rangle }
+{\textstyle{2\over3}}\Theta\omega_\mu +{\textstyle{1\over2}}\curl
A_\mu -\sigma_{\mu\nu}\omega^\nu=0 \,.\label{pe4}
\end{equation}
Shear propagation:
\begin{equation}
\dot{\sigma}_{\langle \mu\nu \rangle }
+{\textstyle{2\over3}}\Theta\sigma_{\mu\nu}
+E_{\mu\nu}-\D_{\langle \mu}A_{\nu\rangle } +\sigma_{\alpha\langle
\mu}\sigma_{\nu\rangle }{}^\alpha+ \omega_{\langle
\mu}\omega_{\nu\rangle} - A_{\langle \mu}A_{\nu\rangle} ={1\over2}
\left({\widetilde{\kappa}\over\kappa}\right)^4 {\cal
P}_{\mu\nu}\,. \label{pe5}
\end{equation}
Gravito-electric propagation:
\begin{eqnarray}
 && \dot{E}_{\langle \mu\nu \rangle }
+\Theta E_{\mu\nu} -\curl H_{\mu\nu}
+{\textstyle{1\over2}}\kappa^2(\rho+p)\sigma_{\mu\nu}
\nonumber\\&&~{} -2A^\alpha\ep_{\alpha\beta(\mu}H_{\nu)}{}^\beta
-3\sigma_{\alpha\langle \mu}E_{\nu\rangle }{}^\alpha
+\omega^\alpha \ep_{\alpha\beta(\mu}E_{\nu)}{}^\beta
\nonumber\\&&~~{}=
-{1\over12}\left({\widetilde{\kappa}\over\kappa}\right)^4\left[
\left\{{\kappa}^4\rho(\rho+p)+8{\cal U}\right\}
\sigma_{\mu\nu}+6\dot{\cal P}_{\langle \mu\nu \rangle}
+2\Theta{\cal P}_{\mu\nu} +6\D_{\langle\mu}{\cal Q}_{\nu\rangle}
\right.\nonumber\\&&~~~\left.{} +12A_{\langle\mu}{\cal
Q}_{\nu\rangle}+6 \sigma^\alpha{}_{\langle\mu} {\cal
P}_{\nu\rangle\alpha}+6 \omega^\alpha\ep_{\alpha\beta(\mu} {\cal
P}_{\nu)}{}^\beta\right] \,. \label{pe6}
\end{eqnarray}
Gravito-magnetic propagation:
\begin{eqnarray}
 &&\dot{H}_{\langle
\mu\nu \rangle } +\Theta H_{\mu\nu} +\curl E_{\mu\nu}-
3\sigma_{\alpha\langle \mu}H_{\nu\rangle }{}^\alpha +\omega^\alpha
\ep_{\alpha\beta(\mu}H_{\nu)}{}^\beta
+2A^\alpha\ep_{\alpha\beta(\mu}E_{\nu)}{}^\beta \nonumber\\&&~~{}=
{1\over2}\left({\widetilde{\kappa}\over\kappa}\right)^4\left[
\curl{\cal P}_{\mu\nu}-3\omega_{\langle\mu} {\cal Q}_{\nu\rangle}
+\sigma^\alpha{}_{(\mu}\ep_{\nu)\alpha\beta} {\cal Q}^\beta\right]
\,. \label{pe7}
\end{eqnarray}
Vorticity constraint:
\begin{equation}
\D^\mu\omega_\mu -A^\mu\omega_\mu =0\,.\label{pcc1}
\end{equation}
Shear constraint:
\begin{equation}
\D^\nu\sigma_{\mu\nu}-\curl\omega_\mu
-{\textstyle{2\over3}}\D_\mu\Theta +2[\omega,A]_\mu =
-\left({\widetilde{\kappa}\over\kappa}\right)^4 {\cal Q}_\mu
 \,.\label{pcc2}
\end{equation}
Gravito-magnetic constraint:
\begin{equation}
 \curl\sigma_{\mu\nu}+\D_{\langle \mu}\omega_{\nu\rangle  }
 -H_{\mu\nu}+2A_{\langle \mu}
\omega_{\nu\rangle  }=0 \,.\label{pcc3}
\end{equation}
Gravito-electric divergence:
\begin{eqnarray}
 && \D^\nu E_{\mu\nu}
 -{\textstyle{1\over3}}\kappa^2\D_\mu\rho
 -[\sigma,H]_\mu
+3H_{\mu\nu}\omega^\nu \nonumber\\&&~~{}=
{1\over18}\left({\widetilde{\kappa}\over\kappa}\right)^4\left\{
{\kappa}^4 \rho\D_\mu\rho +6\D_\mu{\cal U}-6\Theta{\cal
Q}_\mu-9\D^\nu{\cal P}_{\mu\nu} +9\sigma_{\mu\nu}{\cal
Q}^\nu-27[\omega,{\cal Q}]_\mu\right\} \,.\label{pcc4}
\end{eqnarray}
\newpage
Gravito-magnetic divergence:
\begin{eqnarray}
 &&\D^\nu H_{\mu\nu}
-\kappa^2(\rho+p)\omega_\mu +[\sigma,E]_\mu
 -3E_{\mu\nu}\omega^\nu
\nonumber\\&&~~{}=
{1\over6}\left({\widetilde{\kappa}\over\kappa}\right)^4\left\{
{\kappa}^4 \rho(\rho+ p) \omega_\mu - 3\curl{\cal
Q}_\mu+8 {\cal U} \omega_\mu-3[\sigma,{\cal P}]_\mu-3{\cal
P}_{\mu\nu}\omega^\nu\right\} \,.\label{pcc5}
\end{eqnarray}
Here the covariant tensor commutator is
\[
[W,Z]_\mu =\ep_{\mu\alpha\beta}W^\alpha{}_\gamma
Z^{\beta\gamma}\,.
\]
The standard 4-dimensional general relativity results are regained
by setting all right hand sides to zero in
Eqs.~(\ref{pr})--(\ref{pcc5}).

Together with Eqs. (\ref{pc1})--(\ref{pc2'}), these equations
govern the dynamics of the matter and gravitational fields on the
brane, incorporating both the local (quadratic energy-momentum)
and nonlocal (projected Weyl) effects from the bulk. These effects
give rise to important new driving and source terms in the
propagation and constraint equations. The vorticity propagation
and constraint, and the gravito-magnetic constraint have no
explicit bulk effects, but all other equations do. Local and
nonlocal energy density are driving terms in the expansion
propagation, and note that these are the terms that determine the
gravitational and non-gravitational acceleration transverse to the
brane. The spatial gradients of local and nonlocal energy density
provide sources for the gravito-electric field. The nonlocal
anisotropic stress is a driving term in the propagation of shear
and the gravito-electric/ -magnetic fields, and the nonlocal
energy flux is a source for shear and the gravito-magnetic field.

In general the system of equations is not closed: there is no
evolution equation for the nonlocal anisotropic stress ${\cal
P}_{\mu\nu}$, which carries the tensor modes in perturbed
solutions. If we set ${\cal E}_{\mu\nu}=0$ to close the system,
i.e., if we allow only local matter effects from the fifth
dimension, then, as noted above, the density is forced to be
homogeneous. This is clearly too restrictive. A less restrictive
way of closing the system is to assume that the nonlocal
anisotropic stress vanishes, i.e., ${\cal P}_{\mu\nu}=0$. However,
this rules out tensor modes arising from the free field in the
bulk, and also limits the scalar and vector modes, which will in
general also be present in ${\cal P}_{\mu\nu}$.

\subsection{Linearized equations}

We have derived the exact nonlinear equations that govern
gravitational dynamics on the brane as seen by brane observers.
These equations hold for any geometry of the brane, and they are
fully covariant on the brane. In particular, this means that we
can linearize the equations by taking a suitable limit, and
without starting from a given background solution. In this way we
avoid the need for choosing coordinates, and we deal directly with
covariant physical and geometric quantities, rather than metric
components.

The limiting case of the background Friedmann brane is
characterized by the vanishing of all inhomogeneous and
anisotropic quantities. These quantities are then first-order of
smallness in the linearization scheme, and since they vanish in
the background, they are gauge-invariant \cite{eb}. The standard
general relativity scheme is modified by the additional degrees of
freedom arising from bulk effects. In particular, the generalized
Friedmann equation on the brane is \cite{bdel}
\begin{equation}\label{f}
H^2={\textstyle{1\over3}}\Lambda+{\textstyle{1\over3}}\kappa^2\rho
-{K\over a^2} +{\textstyle{1\over36}}\widetilde{\kappa}^4\rho^2+
{1\over3}\left({\widetilde{\kappa}\over\kappa}\right)^4 {\cal
U}_o\left({a_o\over a}\right)^4\,,
\end{equation}
where $K=0,\pm1$.

Linearization about a Friedmann brane model of the propagation and
constraint equations leads to the reduced system:
\begin{eqnarray}
&&\dot{\Theta}+{\textstyle{1\over3}}\Theta^2 -{\rm D}^\mu
A_\mu+{\textstyle{1\over2}}\kappa^2(\rho + 3p) -\Lambda
=
-{1\over12}\left({\widetilde{\kappa}\over\kappa}\right)^4 \left[
{\kappa}^4 \rho(2\rho+3p) +12{\cal U}\right]\,, \label{prl}\\ &&
\dot{\omega}_{ \mu } +2H\omega_\mu +{\textstyle{1\over2}}\curl
A_\mu =0 \,,\label{pe4l}\\ && \dot{\sigma}_{ \mu\nu }
+2H\sigma_{\mu\nu} +E_{\mu\nu}-\D_{\langle \mu}A_{\nu\rangle }
={1\over2} \left({\widetilde{\kappa}\over\kappa}\right)^4 {\cal
P}_{\mu\nu}\,, \label{pe5l}\\ && \dot{E}_{ \mu\nu  } +3H
E_{\mu\nu} -\curl H_{\mu\nu}
+{\textstyle{1\over2}}\kappa^2(\rho+p)\sigma_{\mu\nu}
\nonumber\\&&~~{}=
-{1\over12}\left({\widetilde{\kappa}\over\kappa}\right)^4\left[
\left\{{\kappa}^4\rho(\rho+p)+8{\cal
U}\right\}\sigma_{\mu\nu}+6\D_{\langle\mu}{\cal Q}_{\nu\rangle}
+6\dot{\cal P}_{\mu\nu}+6H{\cal P}_{\mu\nu} \right] \,,
\label{pe6l}\\ &&\dot{H}_{ \mu\nu
 } +3H H_{\mu\nu} +\curl E_{\mu\nu}
={1\over2}\left({\widetilde{\kappa} \over\kappa}\right)^4
\curl{\cal P}_{\mu\nu} \,, \label{pe7l} \\&&\D^\mu\omega_\mu
=0\,,\label{pcc1l}\\ &&\D^\nu\sigma_{\mu\nu}-\curl\omega_\mu
-{\textstyle{2\over3}}\D_\mu\Theta  =
-\left({\widetilde{\kappa}\over\kappa}\right)^4 {\cal Q}_\mu
 \,,\label{pcc2l}\\ &&
\curl\sigma_{\mu\nu}+\D_{\langle \mu}\omega_{\nu\rangle  }
 -H_{\mu\nu}=0 \,,\label{pcc3l}\\ && \D^\nu E_{\mu\nu}
 -{\textstyle{1\over3}}\kappa^2\D_\mu\rho
=
{1\over18}\left({\widetilde{\kappa}\over\kappa}\right)^4\left[
{\kappa}^4 \rho\D_\mu\rho +6\D_\mu{\cal U}-18H{\cal
Q}_\mu-9\D^\nu{\cal P}_{\mu\nu}\right] \,,\label{pcc4l}\\
 &&\D^\nu H_{\mu\nu}
-\kappa^2(\rho+p)\omega_\mu =
{1\over6}\left({\widetilde{\kappa}\over\kappa}\right)^4\left[
\left\{{\kappa}^4 \rho(\rho+ p)+8{\cal U}\right\} \omega_\mu
- 3\curl{\cal Q}_\mu\right] \,,\label{pcc5l}
\end{eqnarray}

These equations, together with the linearized conservation
equations (\ref{pc1}), (\ref{pc2}), (\ref{lc1'}) and (\ref{lc2'}),
are the basis for a covariant and gauge-invariant description of
perturbations on the brane. The local bulk effects (quadratic
energy-momentum effects) are purely scalar, as is the nonlocal
energy density. The nonlocal energy flux has in general scalar and
vector modes:
\begin{equation}\label{pp}
{\cal Q}_\mu=\D_\mu{\cal Q}+\bar{\cal Q}_\mu\,,
\end{equation}
and scalar, vector and tensor modes enter the nonlocal anisotropic
stress:
\begin{equation}\label{p}
{\cal P}_{\mu\nu}=\D_{\langle\mu}\D_{\nu\rangle}{\cal P}
+\D_{\langle\mu}\bar{\cal P}_{\nu\rangle}+\bar{\cal P}_{\mu\nu}\,.
\end{equation}
In these equations, an overbar denotes a transverse
(divergence-free) quantity:
\[
\D^\mu\bar{\cal Q}_\mu=0=\D^\mu\bar{\cal P}_\mu\,,~~
\D^\nu\bar{\cal P}_{\mu\nu}=0\,.
\]

\section{Nonlinear and perturbative dynamics}

Bulk effects introduce new degrees of freedom into the dynamics on
the brane, subject to the additional nonlocal `conservation'
equations (\ref{pc1'}) and (\ref{pc2'}). Standard results in
general relativity may or may not continue to hold under this
higher-dimensional modification. We now use the conservation,
propagation and constraint equations to generalize some standard
results of 4-dimensional general relativity.

\subsection{CMB isotropy and brane homogeneity}

In standard general relativity, the isotropy of the CMB radiation
has crucial implications for the homogeneity of the universe. If
all fundamental observers after last scattering observe an
isotropic CMB, then it follows from a theorem of Ehlers, Geren and
Sachs \cite{egs} that the universe must have a homogeneous
Friedmann geometry. This has been generalized to the
almost-isotropic case \cite{aegs}, providing a foundation for the
perturbative analysis of CMB anisotropies. The Ehlers-Geren-Sachs
theorem is based on the collisionless Boltzmann equation, and on
the kinematic/dynamic characterization:
\begin{equation}\label{egs}
A_\mu=\omega_\mu=\sigma_{\mu\nu}=0\neq\Theta
~\mbox{ and }~q_\mu=\pi_{\mu\nu}=0
~~\Rightarrow~~\mbox{ Friedmann geometry.}
\end{equation}

Bulk effects do not change the Boltzmann equation, but they do
mean that the Friedmann characterization is no longer in general
true on the brane. While the gravito-magnetic constraint Eq.
(\ref{pcc3}) still leads to $H_{\mu\nu}=0$, the shear propagation
equation (\ref{pe5}) no longer forces $E_{\mu\nu}=0$, because of
the nonlocal term ${\cal P}_{\mu\nu}$, so that the intrinsic
metric need not be conformally flat.

A consistent solution on the brane of the system of (nonlinear)
conservation, propagation and constraint equations can be given as
follows. We take
\[
\D_\mu\rho=\D_\mu\rho_{\rm r}=\D_\mu{\cal U}=\D_\mu\Theta=0\,,
~~{\cal Q}_\mu=0=\D^\nu{\cal P}_{\mu\nu}\,.
\]
Then the system of equations reduces to the consistent set
\begin{eqnarray*}
\dot\rho+\Theta\rho&=&0\,,\\
\dot{\rho}_{\rm r}+{\textstyle{4\over3}}\Theta\rho_{\rm r}&=&0\,,\\
\dot{\cal U}+{\textstyle{4\over3}}\Theta{\cal U}&=&0\,,\\
\dot{\Theta}+{\textstyle{1\over3}}\Theta^2+{\textstyle{1\over2}}
\kappa^2(\rho+2\rho_{\rm r})
-\Lambda &=&
-{1\over12}\left({\widetilde{\kappa}\over\kappa}\right)^4 \left[
{\kappa}^4 (\rho+\rho_{\rm r})(2\rho+3\rho_{\rm r})
 +12{\cal U}\right]\,,\\
\dot{\cal P}_{\mu\nu}+{\textstyle{2\over3}}\Theta
{\cal P}_{\mu\nu}&=&0\,,\\
E_{\mu\nu}&=&{\textstyle{1\over2}}\left({\widetilde{\kappa}\over\kappa}
\right)^4 {\cal P}_{\mu\nu}\,.
\end{eqnarray*}

In general, provided that the propagation equation for ${\cal
P}_{\mu\nu}$ is consistent with the bulk geometry, $E_{\mu\nu}$
need not be zero, so that the brane geometry need not be
Friedmann, although it is asymptotically Friedmann.

Thus it is in principle possible via nonlocal bulk effects that
isotropic CMB does {\em not} force the brane metric to be
Friedmann.

\subsection{Generalized gravitational collapse}

Another important question is how the higher-dimensional bulk
effects modify the picture of gravitational collapse/
singularities, which depends on Raychaudhuri's equation \cite{ee}.

The generalized Raychaudhuri equation (\ref{pr}) governs
gravitational collapse and initial singularity behavior on the
brane. The local energy density and pressure corrections,
\[
{\textstyle{1\over12}}\widetilde{\kappa}^4\rho(2\rho+3p)\,,
\]
further enhance the tendency to collapse, if $2\rho+3p>0$. This
condition will be satisfied in thermal collapse (or time-reversed
expansion), but it is violated during very high-energy inflation
($\rho\gg\lambda$), by Eq. (\ref{inf}), and in that case the local
bulk term acts to further accelerate expansion. This is consistent
with the results given in \cite{mwbh}. Thus {\em local bulk
effects at high energy reinforce the formation of singularities
during thermal collapse, as predicted in general relativity, while
further accelerating expansion during high-energy inflation.}

The nonlocal term,
\[
\left({\widetilde{\kappa}\over\kappa}\right)^4{\cal U}\,,
\]
can act either way, depending on its sign. As shown from Eq.
(\ref{6b}), a negative ${\cal U}$ enhances the localization of the
gravitational field on the brane. In this case, the effect of
${\cal U}$ is to counteract gravitational collapse. A positive
${\cal U}$ acts against localization, and also reinforces the
tendency to collapse.

If higher-dimensional corrections to Einstein's theory tend to
{\em prevent} singularities, then the effective energy density
${\cal U}$ on the brane of the free gravitational field in the
bulk should be {\em negative}. In this case, ${\cal U}$ also acts
to reinforce confinement of the gravitational field to the brane.

\subsection{Cosmological scalar perturbations}

The linearized equations on the brane derived in the previous two
sections encompass scalar, vector and tensor modes. In order to
covariantly (and locally) separate out the scalar modes, we impose
the condition that all perturbative quantities are spatial
gradients of scalars, i.e.,
\[
V_\mu=\D_\mu V\,,~~W_{\mu\nu}= \D_{\langle\mu}\D_{\nu\rangle}W\,.
\]
The identities in Appendix B, the vorticity constraint equation
(\ref{pcc1l}) and the gravito-magnetic constraint equation
(\ref{pcc3l}) then show that
\begin{equation}\label{s1}
\curl V_\mu=0=\curl W_{\mu\nu}\,,~
\D^\nu W_{\mu\nu}={\textstyle{2\over3}}\D^2(
\D_\mu W)\,,~\omega_\mu=0=H_{\mu\nu}\,,
\end{equation}
as in standard general relativity.

If we choose the fundamental 4-velocity $u^\mu$ such that
$\sigma_{\mu\nu}=0$, which is the covariant analogue of the
longitudinal or conformal Newtonian gauge in metric-based
perturbation theory (see \cite{ve,m2} for further discussion),
then the shear propagation equation (\ref{pe5l}) becomes a
constraint determining the brane tidal tensor:
\[
E_{\mu\nu}= \D_{\langle\mu}\D_{\nu\rangle}\left[\Phi+
{1\over2}
\left({\widetilde{\kappa}\over\kappa}\right)^4{\cal P}\right]\,.
\]
Here $\Phi$ is the relativistic generalization of the
gravitational potential, defined by $A_\mu=\D_\mu\Phi$, and ${\cal
P}$ is the potential for the nonlocal anisotropic stress, defined
in Eq. (\ref{p}). It follows that {\em nonlocal bulk effects lead
to a change in the gravitational tidal potential} (in
longitudinal-like gauge):
\begin{equation}\label{s2}
\Phi\longrightarrow\Phi+
{1\over2}
\left({\widetilde{\kappa}\over\kappa}\right)^4{\cal P}\,.
\end{equation}
In the general case, i.e., when $u^\mu$ is not chosen to give
$\sigma_{\mu\nu}=0$, this simple relation does not hold.

In order to derive the equations governing density perturbations
in the general case, we define the density and expansion gradients
(as in \cite{eb})
\begin{equation}\label{s3}
\Delta_\mu={a\over\rho}\D_\mu\rho\,,~~Z_\mu=a\D_\mu\Theta\,,
\end{equation}
and the (dimensionless) gradients describing inhomogeneity in the
nonlocal quantities:
\begin{equation}\label{s4}
U_\mu={a\over\rho}\D_\mu{\cal U}\,,~Q_\mu={1\over\rho}
\D_\mu {\cal Q}\,,~P_\mu={1\over a\rho}\D_\mu{\cal P}\,,
\end{equation}
where ${\cal Q}$ is defined in Eq. (\ref{pp}). Then we take the
spatial gradient of the energy conservation equations (\ref{pc1})
and (\ref{lc1'}) and the generalized Raychaudhuri equation
(\ref{prl}), using the identities in Appendix B and the adiabatic
equation (\ref{ad}). We arrive at the following system of
equations:
\begin{eqnarray}
\dot{\Delta}_\mu &=&3wH\Delta_\mu-(1+w)Z_\mu\,,\label{s5}\\
\dot{Z}_\mu &=&-2HZ_\mu-\left({c_{\rm s}^2\over 1+w}\right)
\D^2\Delta_\mu-\left({\widetilde{\kappa}\over\kappa}\right)^4 \rho
U_\mu\nonumber\\ &&~~{}-{\textstyle{1\over2}}\rho\left[\kappa^2+
{\textstyle{1\over6}}\widetilde{\kappa}^4(4+3w)\rho-
\left({\widetilde{\kappa}\over\kappa}\right)^4 \left({2c_{\rm
s}^2\over 1+w}\right){{\cal U}\over\rho}\right]
\Delta_\mu\,,\label{s6}\\ \dot{U}_\mu &=&(3w-1)HU_\mu -
\left({4c_{\rm s}^2\over 1+w}\right)\left({{\cal
U}\over\rho}\right) H\Delta_\mu -\left({4{\cal
U}\over3\rho}\right) Z_\mu-a\D^2Q_\mu\,,\label{s7}\\ \dot{Q}_\mu
&=&(1-3w)HQ_\mu-{1\over3a}U_\mu-{\textstyle{2\over3}}
a\D^2P_\mu+{1\over6a}\left[ \left({8c_{\rm s}^2\over
1+w}\right){{\cal U}\over\rho}-\kappa^4
\rho^2(1+w)\right]\Delta_\mu\,,\label{s8}
\end{eqnarray}
where $w=p/\rho$. The background Friedmann equation (\ref{f})
relates $H$ and $\rho$, with ${\cal U}$ in the background given by
Eq. (\ref{dr}).

In standard general relativity, only the first two equations
apply, with $\widetilde{\kappa}$ set to zero in Eq. (\ref{s6}). In
this case we can decouple the density perturbations via a
second-order equation for $\Delta_\mu$, whose independent
solutions are adiabatic growing and decaying modes. Nonlocal bulk
effects introduce important changes to this simple picture. First,
we note that there is no equation for $\dot{P}_\mu$, so that {\em
in general, scalar perturbations on the brane cannot be predicted
by brane observers without additional information from the
unobservable bulk.}

However, there is a very important exception to this, arising from
the fact that $Q_\mu$ and $P_\mu$ only occur in Eqs.
(\ref{s5})--(\ref{s7}) via the Laplacian terms $\D^2Q_\mu$ and
$\D^2P_\mu$, and the latter term is the only occurrence of $P_\mu$
in the system. It follows that {\em on super-Hubble scales, the
system does close, and brane observers can predict scalar
perturbations from initial conditions intrinsic to the brane.} The
system reduces to 3 coupled equations in $\Delta_\mu$, $Z_\mu$ and
$U_\mu$, plus an equation for $Q_\mu$, which is determined once
the other 3 quantities are solved for. {\em One can decouple the
density perturbations via a third-order equation for
$\Delta_\mu$.} The nonlocal energy density plays the role of a
non-interacting radiation fluid with the same velocity as the
ordinary fluid, and inhomogeneity in ${\cal U}$ introduces an
additional entropy-like scalar mode. As may have been expected,
this additional mode is absent during radiation-domination; in
this case Eqs. (\ref{s5}) and (\ref{s7}) show that
\[
\dot{U}_\mu={{\cal U}_o\over\rho_o}\,\dot{\Delta}_\mu\,.
\]

In principle, it is straightforward to solve the coupled system on
super-Hubble scales, although numerical techniques will be
necessary. We do not attempt particular solutions here. However,
it may be instructive to see the decoupled third-order equation
for density perturbations during matter-domination, on a flat
background:
\begin{equation}
\ddot{\Delta}^{\!\!^{\displaystyle\cdot}}_\mu+2H\ddot{\Delta}_\mu
+\left[{4\over3}\left({\widetilde{\kappa}\over\kappa}\right)^4{\cal U}
-{\textstyle{7\over6}}\kappa^2\rho-{\textstyle{2\over9}}\widetilde
{\kappa}^4\rho^2-{\textstyle{8\over3}}\Lambda\right]\dot{\Delta}_\mu
+\left[{\textstyle{1\over2}}\kappa^2-{\textstyle{2\over3}}
\widetilde{\kappa}^4\rho\right]\rho H\Delta_\mu=0\,. \label{s9}
\end{equation}

\subsection{Cosmological tensor perturbations}

Tensor perturbations are covariantly characterized by
\[
\D_\mu f=0\,,~ A_\mu=\omega_\mu={\cal Q}_\mu=0\,,~ \D^\nu
W_{\mu\nu}=0\,,
\]
where $f=\rho, p, \Theta, {\cal U}$, and
$W_{\mu\nu}=\sigma_{\mu\nu}, E_{\mu\nu}, H_{\mu\nu}, {\cal
P}_{\mu\nu}$. Then all the conservation equations reduce to
background equations, and the system of linearized propagation and
constraint equations in the previous section reduces to:
\begin{eqnarray}
 && \dot{\bar{\sigma}}_{\mu\nu} +2H\bar{\sigma}_{\mu\nu}
+E^*_{\mu\nu}=0\,, \label{pe5t}\\ && \dot{E}^*_{\mu\nu } +3H
E^*_{\mu\nu} -\curl \bar{H}_{\mu\nu}
+{\textstyle{1\over2}}\kappa^2(\rho+p)\bar{\sigma}_{\mu\nu}
\nonumber\\&&~~{}=
-{1\over12}\left({\widetilde{\kappa}\over\kappa}\right)^4\left[
\left\{{\kappa}^4\rho(\rho+p)+8{\cal U}\right\}
\bar{\sigma}_{\mu\nu} +12\left(\dot{\bar{\cal P}}_{ \mu\nu }+2H
\bar{\cal P}_{\mu\nu}\right) \right] \,, \label{pe6t}\\
&&\dot{\bar{H}}_{\mu\nu } +3H \bar{H}_{\mu\nu} +\curl E^*_{\mu\nu}
=0 \,, \label{pe7t} \\&& \curl\bar{\sigma}_{\mu\nu}
 -\bar{H}_{\mu\nu}=0 \,,\label{pcc3t}
\end{eqnarray}
where the overbars denote transverse tensors, and
\[
E^*_{\mu\nu}=\bar{E}_{\mu\nu}-{1\over2}
\left({\widetilde{\kappa}\over\kappa}\right)^4
\bar{\cal P}_{\mu\nu}\,.
\]
Since there is no equation for $\dot{\bar{\cal P}}_{ \mu\nu }$,
the system of equations does not close on the brane: {\em brane
observers cannot evaluate tensor perturbations on the brane
without additional information from the unobservable bulk.} This
remains true on super-Hubble scales, unlike the scalar
perturbation case.

Equations (\ref{pe5t}) and (\ref{pcc3t}) show that the shear is a
gravito-potential for $E^*_{\mu\nu}$ and $\bar{H}_{\mu\nu}$. Using
the identities in Appendix B, we can derive the following
covariant wave equation for the shear:
\begin{eqnarray}
&& \D^2{\bar{\sigma}}_{\mu\nu}-\ddot{\bar{\sigma}}_{\mu\nu}
-5H\dot{\bar{\sigma}}_{\mu\nu}-\left[2\Lambda+{\textstyle{1\over2}}
\kappa^2(\rho-3p)-{\textstyle{1\over12}}\widetilde{\kappa}^4\rho
(\rho+3p)\right]{\bar{\sigma}}_{\mu\nu}\nonumber\\ &&~~{}=-
\left({\widetilde{\kappa}\over\kappa}\right)^4\left[
\dot{\bar{\cal P}}_{ \mu\nu }+2H \bar{\cal P}_{\mu\nu} \right] \,.
\label{t1}
\end{eqnarray}
For adiabatic tensor perturbations in standard general relativity,
the right hand side falls away. Nonlocal bulk effects provide
driving terms that are like anisotropic stress terms in general
relativity \cite{c}. In the latter case however, the evolution of
anisotropic stress is determined by the Boltzmann equation or
other intrinsic physics.

What we can conclude from Eq. (\ref{t1}) is that the {\em local
bulk effects enhance tensor perturbations for non-inflationary
expansion, and suppress them during inflation}, since inflation
implies $\rho+3p<0$ (at all energy scales) by Eq. (\ref{inf}). The
nature of the {\em nonlocal} bulk effects carried by $\bar{\cal
P}_{\mu\nu}$ requires knowledge of the off-brane dynamics of
${\cal E}_{AB}$, which we have not considered.

\subsection{Cosmological vector perturbations}

The linearized vorticity propagation equation (\ref{pe4l}) does
not carry any bulk effects, and vorticity decays as in standard
general relativity. However, the gravito-magnetic divergence
equation (\ref{pcc5l}) shows that the local and nonlocal energy
density and the nonlocal energy flux provide additional sources
for the brane gravito-magnetic field:
\[
\D^\nu H_{\mu\nu}
=\kappa^2(\rho+p)\omega_\mu +
{1\over6}\left({\widetilde{\kappa}\over\kappa}\right)^4\left[
\left\{{\kappa}^4 \rho(\rho+ p)+8{\cal U}\right\} \omega_\mu
- 3\curl{\cal Q}_\mu\right] \,.
\]
In standard general relativity, it is necessary to increase the
angular momentum density $\kappa^2(\rho+p)\omega_\mu$ in order to
increase the gravito-magnetic field, but bulk effects allow an
increased gravito-magnetic field without this. In particular,
unlike in general relativity, {\em it is possible to source vector
perturbations even when the vorticity vanishes,} since $\curl{\cal
Q}_\mu$ may be nonzero.

\newpage
\section{Conclusion}

By adopting a covariant approach based on physical and geometrical
quantities that are in principle measurable by brane-observers, we
have given a comprehensive analysis of intrinsic cosmological
dynamics in Randall-Sundrum-type brane-worlds. This has allowed us
to carefully delineate what can and can not be predicted by brane
observers without additional information from the unobservable
bulk.

Our main result is probably that {\em scalar perturbations on
super-Hubble scales can be evaluated intrinsically on the brane.}
This is a generalization of the result given in \cite{mwbh}, i.e.,
the evaluation of adiabatic scalar perturbations on super-Hubble
scales during inflation on the brane, done under the assumption
that ${\cal E}_{\mu\nu}$ may be neglected.

We showed that tensor perturbations can not be evaluated
intrinsically on any scales. This is not surprising, since the
ability of gravitational waves to penetrate the 5th dimension
inevitably introduces unpredictability from the standpoint of
observers confined to the brane. Vector perturbations on the brane
can be generated even in the absence of vorticity, via the curl of
the nonlocal energy flux.

Our nonperturbative results include:\\ (i) Calculating the
gravitational and non-gravitational acceleration felt by brane
observers, allowing us to provide covariant characterizations of
gravity and matter localization, and showing in particular that
the non-gravitational off-brane acceleration of fundamental
world-lines is towards the brane during inflation. \\ (ii) Showing
how bulk effects can disrupt the relation between isotropy of the
CMB and spatial isotropy and homogeneity on the brane.\\ (iii)
Showing how bulk effects modify the dynamics of gravitational
collapse and singularity formation.

We have derived the exact nonlinear equations governing
cosmological dynamics on the brane, and the corresponding
covariant and gauge-invariant linearized equations. Further
implications of these equations could usefully be pursued. In
particular, an important topic for further research is the
calculation of scalar perturbations on very large scales, and the
investigation of limits imposed by observations.

However, the major further step required, and not undertaken here,
is to complete the picture by investigating the dynamical
equations of the gravitational field off the brane. A starting
point is provided by the general equations given in
\cite{sms,ssm}, which determine
\[
{\cal L}_{\bf n}{\cal E}_{AB}\,,~
{\cal L}_{\bf n}{\cal B}_{ABC}\,,~
{\cal L}_{\bf n}{R}_{ABCD}\,,
\]
where $R_{ABCD}$ is the 4-dimensional Riemann tensor, and
\[
{\cal B}_{ABC}=g_A{}^D g_B{}^E\widetilde{C}_{DECF}n^F\,.
\]
However, it may turn out to be more useful to develop an
alternative decomposition of the bulk Weyl tensor, along a
timelike direction $\widetilde{u}^A$ rather than a spatial
direction $n^A$. Intuitively, this may provide a more direct and
transparent route to the evolution equation for ${\cal
P}_{\mu\nu}$, whose absence leads to unpredictability on the
brane. Such a timelike decomposition requires a generalization to
higher dimensions of the 4-dimensional decomposition of the Weyl
tensor \cite{ee}. The generalization has been given in \cite{s}.

The complete and closed system of dynamical equations would allow
us to develop more systematic and probing tests of the
Randall-Sundrum-type models against observational constraints.
Despite the appealing geometric and particle-physics properties of
such models, it is their confrontation with cosmological
observational tests that will provide a more decisive arbiter as
to whether they are viable generalizations of Einstein's theory.
\[ \]
{\em Note:} Since this work was completed, a number of papers has
appeared, setting up the 5-dimensional formalism (metric-based) of
bulk perturbations~\cite{new}, and gravitational waves produced
during inflation on the brane have also been studied~\cite{lmw}.

\vfill {\noindent{\bf Acknowledgements:}\\

I would like to thank Bruce Bassett, Marco Bruni, Chris Clarkson,
Naresh Dadhich, George Ellis, Jose Senovilla and David Wands for
helpful discussions and comments.}

\newpage
\appendix

\section{General propagation and constraint equations}

For a general, imperfect energy-momentum tensor, as in Eq.
(\ref{3''}), the propagation and constraint equations
(\ref{pr})--(\ref{pcc5}) are generalized to:\\

\noindent{\sf Propagation:}
\begin{eqnarray}
&&\dot{\Theta}+{\textstyle{1\over3}}\Theta^2+
\sigma_{\mu\nu}\sigma^{\mu\nu} -2\omega_\mu\omega^\mu -{\rm D}^\mu
A_\mu+A_\mu A^\mu+{\textstyle{1\over2}}\kappa^2(\rho + 3p)
-\Lambda \nonumber\\
&&~~{}=-{1\over12}\left({\widetilde{\kappa}\over\kappa}\right)^4
\left[{\kappa}^4\left(2\rho^2+3\rho p-3 q_\mu q^\mu\right)
+12{\cal U}\right]\,, \label{r}\\&&{}\nonumber\\ &&
\dot{\omega}_{\langle \mu\rangle }
+{\textstyle{2\over3}}\Theta\omega_\mu +{\textstyle{1\over2}}\curl
A_\mu -\sigma_{\mu\nu}\omega^\nu=0 \,,\label{e4}\\
&&{}\nonumber\\&& \dot{\sigma}_{\langle \mu\nu \rangle }
+{\textstyle{2\over3}}\Theta\sigma_{\mu\nu}
+E_{\mu\nu}-{\textstyle{1\over2}}\kappa^2\pi_{\mu\nu} -\D_{\langle
\mu}A_{\nu\rangle } +\sigma_{\alpha\langle \mu}\sigma_{\nu\rangle
}{}^\alpha+ \omega_{\langle \mu}\omega_{\nu\rangle} - A_{\langle
\mu}A_{\nu\rangle}\nonumber\\&&~~{}={1\over24}
\left({\widetilde{\kappa}\over\kappa}\right)^4\left[ {\kappa}^4
\left\{-(\rho+3p)\pi_{\mu\nu}+
\pi_{\alpha\langle\mu}\pi_{\nu\rangle}{}^\alpha+q_{\langle\mu}q_
{\nu\rangle}\right\} +12{\cal P}_{\mu\nu}\right] \,, \label{e5}\\
&&{}\nonumber\\&& \dot{E}_{\langle \mu\nu \rangle } +\Theta
E_{\mu\nu} -\curl H_{\mu\nu}
+{\textstyle{1\over2}}\kappa^2(\rho+p)\sigma_{\mu\nu}
+{\textstyle{1\over2}}\kappa^2\dot{\pi}_{\langle \mu\nu\rangle  }
+{\textstyle{1\over2}}\kappa^2\D_{\langle \mu}q_{\nu\rangle }
+{\textstyle{1\over6}}\kappa^2 \Theta\pi_{\mu\nu}
+\kappa^2A_{\langle \mu}q_{\nu\rangle} \nonumber\\ &&~{}
-2A^\alpha\ep_{\alpha\beta(\mu}H_{\nu)}{}^\beta
-3\sigma_{\alpha\langle \mu}E_{\nu\rangle }{}^\alpha
+\omega^\alpha \ep_{\alpha\beta(\mu}E_{\nu)}{}^\beta
+{\textstyle{1\over2}}\kappa^2\sigma^\alpha{}_{\langle
\mu}\pi_{\nu\rangle \alpha} +{\textstyle{1\over2}}\kappa^2
\omega^\alpha\ep_{\alpha\beta(\mu}\pi_{\nu)}{}^\beta
\nonumber\\&&~~{}=
{1\over72}\left({\widetilde{\kappa}\over\kappa}\right)^4\left[
-{\kappa}^4\left\{3 \left( 2\rho^2+2\rho
p-\pi_{\alpha\beta}\pi^{\alpha\beta} -2q_\alpha q^\alpha\right)
\sigma_{\mu\nu}+3(\dot\rho+\dot p)\pi_{\mu\nu} +3(\rho+3p)
\dot{\pi}_{\langle\mu\nu\rangle}
\right.\right.\nonumber\\&&~~~\left.\left.{}-
6\pi_{\alpha\langle\mu} \dot{\pi}_{\nu\rangle}{}^\alpha
-6q_{\langle\mu} \dot{q}_{\nu\rangle}+ {\textstyle{3\over2}}
\D_{\langle\mu}\left( \pi_{\nu\rangle\alpha}q^\alpha-4\rho
q_{\nu\rangle}\right) +\Theta\left([\rho+3p]\pi_{\mu\nu}-
\pi_{\alpha\langle\mu} \pi_{\nu\rangle}{}^\alpha
-q_{\langle\mu}q_{\nu\rangle}\right) \right.\right.
\nonumber\\&&~~~\left.\left.{} - 3A_{\langle\mu}\left(4\rho
q_{\nu\rangle}- \pi_{\nu\rangle\alpha} q^\alpha\right)+
3(\rho+3p)\sigma^\alpha{}_{\langle\mu}\pi_{\nu\rangle\alpha}
-3\sigma^\alpha{}_{\langle\mu}
\left(\pi^\beta{}_{\langle\nu\rangle}\pi_{\alpha\rangle\beta}
+q_{\nu\rangle} q_\alpha\right)
\right.\right.\nonumber\\&&~~~\left.\left.{}
 +3(\rho+3p) \omega^\alpha
\ep_{\alpha\beta(\mu}\pi_{\nu)}{}^\beta -3\omega_\alpha
\ep^{\alpha\beta}{}{}_{(\mu}\left(
\pi^\gamma{}_{\langle\nu)}\pi_{\beta\rangle\gamma} + q_{\nu)}
q_\beta \right) \right\} -48{\cal U}\sigma_{\mu\nu}-36\dot{\cal
P}_{\langle \mu\nu \rangle} \right. \nonumber\\
&&~~~\left.{}-36\D_{\langle\mu}{\cal Q}_{\nu\rangle}
-12\Theta{\cal P}_{\mu\nu} -72A_{\langle\mu}{\cal
Q}_{\nu\rangle}-36 \sigma^\alpha{}_{\langle\mu} {\cal
P}_{\nu\rangle\alpha}- 36 \omega^\alpha\ep_{\alpha\beta(\mu} {\cal
P}_{\nu)}{}^\beta\right] \,, \label{e6}\\
&&{}\nonumber\\&&\dot{H}_{\langle \mu\nu \rangle } +\Theta
H_{\mu\nu} +\curl E_{\mu\nu}
-{\textstyle{1\over2}}\kappa^2\curl\pi_{\mu\nu}-
3\sigma_{\alpha\langle \mu}H_{\nu\rangle }{}^\alpha +\omega^\alpha
\ep_{\alpha\beta(\mu}H_{\nu)}{}^\beta \nonumber\\
&&~{}+2A^\alpha\ep_{\alpha\beta(\mu}E_{\nu)}{}^\beta
+{\textstyle{3\over2}}\kappa^2\omega_{\langle \mu}q_{\nu\rangle }-
{\textstyle{1\over2}}\kappa^2\sigma^\alpha{}_{(\mu}
\ep_{\nu)\alpha\beta}q^\beta \nonumber\\&&~~{}=
{1\over48}\left({\widetilde{\kappa}\over\kappa}\right)^4\left[
{\kappa}^4\left\{ -2\curl\left((\rho+3p)\pi_{\mu\nu}-
\pi_{\alpha\langle\mu}\pi_{\nu\rangle}{}^\alpha - q_{\langle\mu}
q_{\nu\rangle} \right)-12\rho\omega_{\langle\mu} q_{\nu\rangle}+ 4
\omega_{\langle\mu} \pi_{\nu\rangle\alpha}q^\alpha\right.\right.
\nonumber\\ &&~~~\left.\left.{}+ \sigma^\alpha{}_{(\mu}
\ep_{\nu)\alpha\beta} \left(4\rho q^\beta-\pi^{\beta\gamma}
q_\gamma\right)\right\}+24\curl{\cal
P}_{\mu\nu}-72\omega_{\langle\mu} {\cal Q}_{\nu\rangle}
+24\sigma^\alpha{}_{(\mu}\ep_{\nu)\alpha\beta} {\cal
Q}^\beta\right] \,. \label{e7}
\end{eqnarray}\\

\noindent{\sf Constraint:}
\begin{eqnarray}
&&\D^\mu\omega_\mu -A^\mu\omega_\mu =0\,,\label{cc1}\\
&&{}\nonumber\\&&\D^\nu\sigma_{\mu\nu}-\curl\omega_\mu
-{\textstyle{2\over3}}\D_\mu\Theta +\kappa^2q_\mu
+2[\omega,A]_\mu
=
-{1\over24}\left({\widetilde{\kappa}\over\kappa}\right)^4\left[
{\kappa}^4\left(4\rho q_\mu-\pi_{\mu\nu}q^\nu\right)
+24{\cal Q}_\mu\right]
 \,,\label{cc2}\\&&{}\nonumber\\ &&
\curl\sigma_{\mu\nu}+\D_{\langle \mu}\omega_{\nu\rangle  }
 -H_{\mu\nu}+2A_{\langle \mu}
\omega_{\nu\rangle  }=0 \,,\label{cc3}\\&&{}\nonumber
\\ && \D^\nu E_{\mu\nu}
+{\textstyle{1\over2}}\kappa^2\D^\nu\pi_{\mu\nu}
 -{\textstyle{1\over3}}\kappa^2\D_\mu\rho
+{\textstyle{1\over3}}\kappa^2\Theta q_\mu -[\sigma,H]_\mu
+3H_{\mu\nu}\omega^\nu
-{\textstyle{1\over2}}\kappa^2\sigma_{\mu\nu}q^\nu+
{\textstyle{3\over2}} \kappa^2[\omega,q]_\mu \nonumber\\&&~~{}=
{1\over48}\left({\widetilde{\kappa}\over\kappa}\right)^4\left[
{\kappa}^4\left\{{\textstyle{2\over3}}\Theta\left(
\pi_{\mu\nu}q^\nu-4\rho q_\mu\right) +2\D^\nu\left((\rho+3p)
\pi_{\mu\nu} -\pi_{\alpha\langle\mu} \pi_{\nu\rangle}{}^\alpha-
q_{\langle\mu} q_{\nu\rangle}\right)\right.\right.\nonumber\\
&&~~~{}\left.\left.+ {\textstyle{8\over3}}\rho\D_\mu\rho
-4\pi^{\alpha\beta}\D_\mu\pi_{\alpha\beta}
+\sigma_{\mu\nu}\left(4\rho q^\nu-\pi^{\nu\alpha}
q_\alpha\right)+3\ep_{\mu\alpha\beta}
\omega^\alpha\pi^{\beta\gamma}
q_\gamma-4\rho[\omega,q]_\mu\right\}\right.\nonumber\\
&&~~~{}\left. +16\D_\mu{\cal U}-16\Theta{\cal Q}_\mu-24\D^\nu{\cal
P}_{\mu\nu} +24\sigma_{\mu\nu}{\cal Q}^\nu-72[\omega,{\cal
Q}]_\mu\right] \,,\label{cc4}\\ &&{}\nonumber\\ &&\D^\nu
H_{\mu\nu} +{\textstyle{1\over2}}\kappa^2\curl q_\mu
-\kappa^2(\rho+p)\omega_\mu +[\sigma,E]_\mu+{\textstyle{1\over2}}
\kappa^2 [\sigma,\pi]_\mu -3E_{\mu\nu}\omega^\nu
+{\textstyle{1\over2}}\kappa^2\pi_{\mu\nu} \omega^\nu
\nonumber\\&&~~{}=
{1\over48}\left({\widetilde{\kappa}\over\kappa}\right)^4\left[
{\kappa}^4\left\{\curl\left(\pi_{\mu\nu}q^\nu-4\rho q_\mu
\right)+4\left(2\rho^2+2\rho p-\pi_{\alpha\beta}\pi^{\alpha\beta}
-2q_\alpha q^\alpha\right)\omega_\mu
\right.\right.\nonumber\\&&\left.\left.~~~{}
-2\ep_{\mu\alpha\beta}
\sigma^\alpha{}_\gamma\left(\pi_\nu{}^{\langle\beta}
\pi^{\gamma\rangle\nu} +q^{\langle\beta}q^{\gamma\rangle}\right)
+2\left((\rho+3p)\pi_{\mu\nu}
-\pi_{\alpha\langle\mu}\pi_{\nu\rangle}{}^\alpha- q_{\langle\mu}
q_{\nu\rangle}\right)\omega^\nu
\right.\right.\nonumber\\&&\left.\left.~~~{}
+2(\rho+3p)[\sigma,\pi]_\mu\right\} - 24\curl{\cal Q}_\mu+64 {\cal
U} \omega_\mu-24[\sigma,{\cal P}]_\mu-24{\cal
P}_{\mu\nu}\omega^\nu\right] \,.\label{cc5}
\end{eqnarray}\\

The 4-dimensional general relativistic results are regained by
setting all the right hand sides of these equations to 0.

\section{Differential identities}

On a flat Friedmann background, the following covariant linearized
identities hold \cite{m2}:
\begin{eqnarray}
\D_\mu \dot f &=& (\D_\mu f)^{\displaystyle\cdot}+H\D_\mu f- \dot
f A_\mu\,,\\ \D^2(\D_\mu f) &=& \D_\mu(\D^2 f)+2\dot
f\omega_\mu\,,\\ (\D^2f)^{\displaystyle\cdot}& =&\D^2\dot
f-2H\D^2f+\dot f\D^\mu A_\mu\,,\\ \curl\D_\mu f &=&-2\dot
f\omega_\mu\,,\\ \curl \D_{\langle\mu}\D_{\nu\rangle}f &=&0\,,\\
(\D_\mu V_\nu)^{\displaystyle\cdot}&=&\D_\mu\dot{V}_\nu-H \D_\mu
V_\nu\,,\\ \D_{[\mu}\D_{\nu]}V_\alpha
&=&0=\D_{[\mu}\D_{\nu]}W_{\alpha\beta}\,,\\ \D^\mu\curl V_\mu
&=&0\,,\\ \D^\nu\D_{\langle\mu}V_{\nu\rangle}&=&
{\textstyle{1\over2}}\D^2V_\mu +{\textstyle{1\over6}}
\D_\mu(\D^\nu V_\nu)\,,\\
\curl\D_{\langle\mu}V_{\nu\rangle}&=&{\textstyle{1\over2}}
\D_{\langle\mu}\curl V_{\nu\rangle}\,,\\ \curl\curl V_\mu &=&-\D^2
V_\mu+\D_\mu(\D^\nu V_\nu)\,,\\ (\D_\mu
W_{\alpha\beta})^{\displaystyle\cdot}&=&\D_\mu
\dot{W}_{\alpha\beta}-H\D_\mu W_{\alpha\beta}\,,\\ \D^\nu\curl
W_{\mu\nu} &=&{\textstyle{1\over2}}\curl( \D^\nu W_{\mu\nu})\,,\\
\curl\curl W_{\mu\nu} &=&-\D^2 W_{\mu\nu}
+{\textstyle{3\over2}}\D_{\langle\mu}\D^\alpha
W_{\nu\rangle\alpha}\,,
\end{eqnarray}
where $V_\mu=V_{\langle\mu\rangle}$
and $W_{\mu\nu}=W_{\langle\mu\nu\rangle}$ vanish in the
background.

\end{document}